# RACE: A Reinforcement Learning Framework for Improved Adaptive Control of NoC Channel Buffers

Kamil Khan
Electrical and Computer Engineering
Colorado State University
Fort Collins, CO, USA
kamil@colostate.edu

Sudeep Pasricha
Electrical and Computer Engineering
Colorado State University
Fort Collins, CO, USA
sudeep@colostate.edu

Ryan Gary Kim
Electrical and Computer Engineering
Colorado State University
Fort Collins, CO, USA
ryan.g.kim@colostate.edu

## ABSTRACT

Network-on-chip (NoC) architectures rely on buffers to store flits to cope with contention for router resources during packet switching. Recently, reversible multi-function channel (RMC) buffers have been proposed to simultaneously reduce power and enable adaptive NoC buffering between adjacent routers. While adaptive buffering can improve NoC performance by maximizing buffer utilization, controlling the RMC buffer allocations requires a congestion-aware, scalable, and proactive policy. In this work, we present RACE, a novel reinforcement learning (RL) framework that utilizes better awareness of network congestion and a new reward metric ("falsefulls") to help guide the RL agent towards better RMC buffer control decisions. We show that RACE reduces NoC latency by up to 48.9%, and energy consumption by up to 47.1% against state-of-the-art NoC buffer control policies.

## 1 INTRODUCTION

As the number of cores in a single chip continue to scale, network-on-chips (NoCs) have emerged as the leading communication solution in manycore systems [1, 2, 3]. By using routers to forward messages toward their destination, router buffers to hold messages that cannot be immediately forwarded, and links to connect these routers, a NoC provides a scalable, high-performance interconnect fabric between cores [18]. However, NoC architectures can contribute significantly to the chip's total power. For example, the NoC contributed 28% and 19% of the tile power for Intel's 80-core TeraFlop [1] and MIT's 36-core Scorpio [2] chips, respectively.

In particular, the router buffers consume a significant amount of the power and latency in the NoC [3]. To combat this, prior work has proposed to convert some/all router buffers to channel buffers (i.e., buffers along inter-router links) [4] or eliminate buffers entirely [5, 6]. In [5], a bufferless router and routing scheme reduced NoC energy by 42.5% on average. On the other hand, [4, 7] moved buffers from the router and onto the channel by replacing some channel repeaters with reversible multi-function channel (RMC) buffers. In such RMC-based architectures, each inter-router channel contains RMC buffers grouped into multiple subchannels that can be configured to store or propagate flits in either direction. By sharing logic for both channel repeaters and buffering, [4, 7] reduced the router power by 24%. In addition to the power saving benefits of RMC-based systems, these reversible subchannels also present an opportunity to dynamically allocate buffers to the routers on either side of the channel. This can help adjust the buffers to cater to spatiotemporal changes in the traffic demands. By using this capability, both [6] and [7] were able to reduce the latency and increase the saturation throughput.

In this paper, we similarly consider an RMC-based NoC system where many of the router buffers are moved to the channel and develop a high performance RMC subchannel control policy that outperforms state-of-the-art approaches for RMC control. Our novel contributions in this work are as follows:

- We motivate the need for a novel proactive per-RMC approach to controlling reversible NoC channel buffers.
- We devise a new reinforcement learning based RMC control policy and framework, RACE, that uses deep Q-learning to proactively cater to changing NoC buffer demands.
- We propose a novel state definition that helps estimate regional NoC congestion and a new reward metric that is strongly aligned with RMC control goals, as part of RACE.

RACE uses reinforcement learning to proactively adjust RMC subchannel directions in NoCs when the demand for buffering changes, with changing application traffic phases. Here, we place particular emphasis on providing a scalable technique that can improve network performance. We demonstrate that RACE is able to significantly outperform state-of-the-art heuristic and RL-based NoC buffer control policies for RMC-based systems.

## 2 RELATED WORK

Due to the cost of buffers in NoCs, many works have examined improving the efficiency and utilization of these buffers. Channel buffers were introduced to reduce router power consumption [4] by replacing some of the router buffers with modified low-power tri-state repeaters in the channel. By using fewer total transistors while reusing the existing repeater circuit, channel buffers reduce total NoC power significantly [4]. These channel buffers store flits when the receiving router buffer is full. However, the channel buffers still suffer from under-utilization as traffic varies.

To help combat non-uniform, dynamic buffer demand, static design-time buffer sizing and dynamic rearrangement of router buffers have been proposed [8]. While design-time buffer sizing considers the spatial variance in buffer demand, it cannot adapt to temporal variations. Alternatively, [9] proposed a reversible link architecture that divided flits and links into smaller "phits". The links can then configure some number of phit links in either direction to adapt to dynamic bandwidth demand. Unfortunately, due to the serialization and deserialization of flits into phits, this resulted in a complicated router architecture.

By combining channel buffers and reversibility, [6] created the reversible multi-function channel (RMC) to improve fault tolerance, reduce power, and provide adaptive bandwidth. These RMC buffers (see Fig. 1(a)) replace some of the repeaters and can store or propagate a flit in either direction. By reconfiguring the RMCs, we can dynamically allocate channel buffers to either direction to handle both spatial and temporal demand variance.

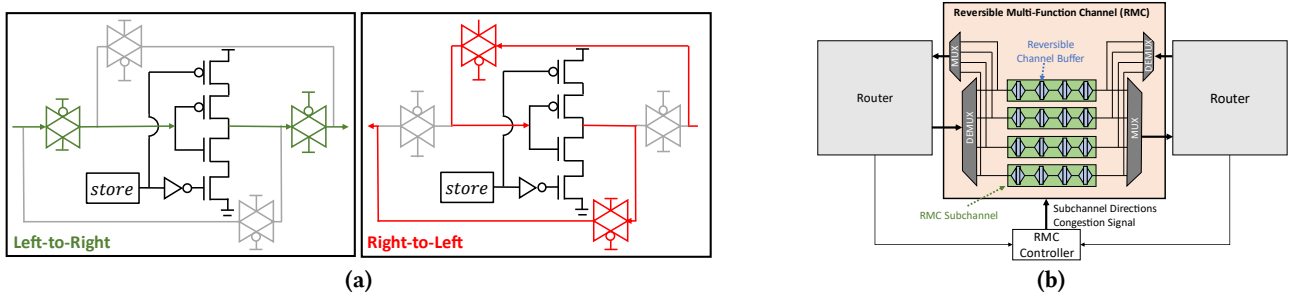

**Figure 1.** (a) Circuit diagram of a single reversible channel buffer when configured in each direction. (b) The reversible multi-function channel (RMC) architecture along with the RMC controller [7].

To control the RMC directions, heuristics [6] and machine learning techniques such as decision-trees [10] and reinforcement learning (RL) [7, 12] have been proposed. In QORE [6], more RMC buffers were allocated in a particular direction if the difference in flit traversals was greater than 5% for that direction over the last 50-cycle reconfiguration interval (epoch). However, we have found that buffer demand rarely stays constant across epochs, leading to sub-optimal configurations. In CURE [7], the authors proposed an RL-based per-router controller that selects the RMC configuration with fault and power management decisions. In their approach, the state contained only local router information while the action assigned the same configuration to all RMCs connected to the router. As we will demonstrate in our motivation (Section 3) and results (Section 5), this RL formulation leads to many suboptimal decisions in an NoC architecture.

In this work, we improve upon the RMC architecture from previous work [7] and present a novel per-RMC RL-based policy. We utilize the flow control credits in adjacent routers to generate a better estimate of the network congestion in a NoC region, leading to better decision making. We also develop a new reward metric that better represents our RMC buffer control goals, resulting in significantly lower latency and energy overhead than the state-of-the-art RL techniques for NoC buffer control.

## 3 BACKGROUND AND MOTIVATION

### 3.1 Reversible Multi-Function Channel (RMC)

In this paper, we consider the reversible multi-function channel (RMC) inter-router link design from CURE [7] (see Fig. 1(b)). In this case, the two unidirectional inter-router links in traditional NoCs are replaced with a single RMC consisting of multiple reversible physical subchannels. Each of the subchannels have some of the traditional repeaters replaced with RMC buffers. In Fig. 1(b), four subchannels and four RMC buffers per subchannel are shown as an example. Each subchannel is 128-bits wide, as advocated in [7], although theoretically it can assume any value.

The RMC buffers get their name from having the ability to switch between storage and repeater (forwarding) functions, and the ability to switch directions. When a router's input buffer is full, the router will send a congestion signal to the RMC controller that will switch the RMC buffers to storage mode. As the congestion clears, flits are moved into the router from the RMC buffers. Although there are multiple physical subchannels, only a single flit can be transmitted in each direction per cycle. If more buffers are needed in a direction, some of the subchannel's directions can be reversed. However, a subchannel cannot be reversed immediately if it is not empty. In this case, the controller will disable write functionality and wait for all subchannel buffers to be emptied into the router and then change the direction of the subchannel.

### 3.2 Opportunity for Dynamic Buffering

In an NoC, input router buffers are used to store flits that cannot be immediately forwarded. This allows flits later in the packet to move out of the way even if there is congestion at the head-of-line, thus reducing delays [17]. The number of input buffers are typically fixed uniformly per port at design time. While such a design offers simplicity, it is inefficient due to the spatiotemporal variations typically seen in the buffer demands. For example, during a memory-intensive application phase, we would expect to see much higher traffic near last-level caches and memory controllers, resulting in higher buffer demands in those areas of the NoC. Similarly, compute-intensive application phases or idle cores have lower buffer demands in the connecting router.

These examples exist on the extreme ends of the buffer demand spectrum. Unfortunately, the actual buffer demand typically exists somewhere in between these extremes and is hard to predict due to complex network effects. In Fig. 2, the average buffer utilization associated with different input ports near the center of an 8x8 NoC are shown for three consecutive 50-cycle time windows. Here, the NoC is handling uniform random traffic. The buffer utilizations vary significantly across the three time windows. In the first time window, the highest buffer utilization belongs to the buffer going from router 35 to 34. Then, it switches to the buffer going from router 35 to 27 in the second time window and then router 35 to 43 in the last time window. Using the reversing capabilities of the RMC, we could reverse some of the RMC buffers going from router 27 to 35 in the second time window, allocating more buffers in the 35 to 27 direction. With timely and harmonious reversals of the RMC subchannels, we can maximally utilize the NoC buffers and improve performance and energy. Based on our RTL-level implementation and analysis, the reversal latency (time required to complete a reversal) can vary from 5-11 clock cycles per reconfiguration depending on traffic conditions and the RMC control policy used. This overhead can degrade performance significantly if subchannels are reconfigured frequently, motivating the need for intelligent reconfiguration, as discussed next.

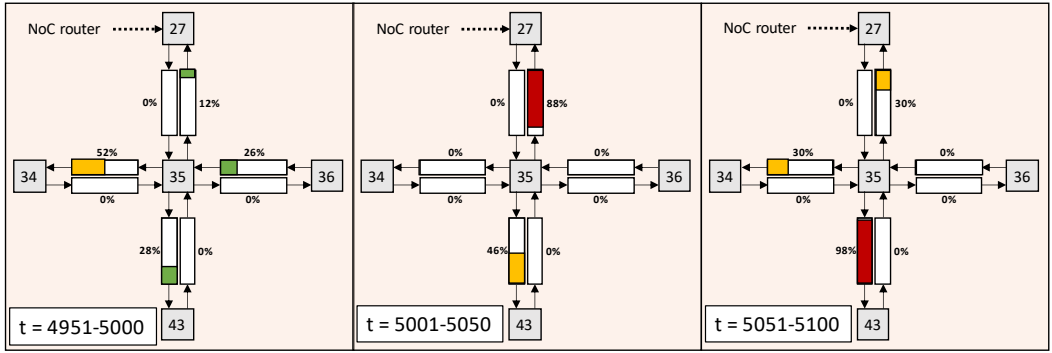

**Figure 2. Average RMC buffer utilization on the links connected to router 35 in an 8x8 mesh system (near the middle) for three consecutive 50-cycle time windows.**

### 3.3 Motivation for Reinforcement Learning

To allocate the RMC buffers, one option is a simple policy that allocates the subchannels based on historical information about buffer utilization in each direction on a link. If one direction was observed to be sending many more flits, it would be more likely that it could take advantage of more buffers. The simple policy presented in [6] is based on this concept. However, there are a few problems with the policy in [6]:

(1) This policy relies on having similar buffer demands from one epoch to the next by using historical bandwidth information. For this approach to work, buffer demand needs to remain steady which is not the case as shown in Fig. 2 and also in real application traffic scenarios. In any case, the number of flit traversals does not always correlate with buffer demand, especially during periods of low congestion.
(2) This policy only examines the local RMC, which can lead to problems in inferring buffer demand since congestion is a regional issue and depends on the rest of the network.
(3) As the subchannel cannot reverse immediately if it has flits (see Section 3.2), a policy needs to change the configuration early enough so that the flits have time to leave the RMC. This policy cannot proactively adjust to consider this issue.

As an example, in Fig. 2, this simple policy [6] would only allocate equal subchannels in both directions for the RMC between 27 and 35 during the second time window because the previous window had minimal differences in load between each direction. To address these problems, reinforcement learning (RL) presents a potential solution. Here, an RL policy can utilize multiple inputs from across a NoC region to help infer regional congestion (problem (2)) while its proactive nature can help predict future buffer demands (problem (1)) and change the configuration early enough so that the RMC will be ready (problem (3)).

## 4 PROPOSED RMC CONTROL POLICY

To take full advantage of RMC buffers in NoCs, we propose a per-RMC reinforcement learning (RL) policy, RACE, to select the number of subchannels in each direction. Unlike previous RL solutions, e.g., [7] and [11], we take advantage of regional NoC information and create a more relevant reward metric to produce a scalable and high-performing RMC control policy.

### 4.1 Reinforcement Learning Preliminaries

RL has found recent success in robotics, computer vision, natural language processing, protein folding, healthcare, etc. [12]. RL works by enabling agents to learn policies that maximize a "reward" by interacting with their environment [13]. A popular way to have the RL agent learn is through a Q-learning algorithm [14]. Using the received reward at each epoch $t$, we update the quality of taking an action $a$ in state $s$, $Q(s,a)$, using the Q-learning update function:

$$Q(s_t, a_t) \leftarrow Q(s_t, a_t) + \alpha \cdot \left( r_t + \gamma \cdot \max_a Q(s_{t+1}, a) - Q(s_t, a_t) \right) \quad (1)$$

where $r_t$ is the immediate reward received, $\max_a Q(s_{t+1}, a)$ is the maximum Q-value for the next state, $\gamma$ is the discount factor which can be used to adjust the importance of future reward, and $\alpha$ is the learning rate which controls the size of the update.

Traditionally, the Q-function, $Q(s,a)$, is stored using a table with a value for each $(s,a)$ pair. However, the table size grows quickly as the number of possible states and actions increases, resulting in large area and energy overheads. Consequently, we use Deep Q-Learning (DQL) [13], which utilizes artificial neural networks (ANN) to train and represent $Q(s,a)$, thus reducing hardware and performance overhead.

### 4.2 State, Action, and Reward Definition

Our per-RMC RL agents attempt to determine the best number of subchannels in each direction in every epoch (a fixed interval of time; see Section 5.2 for the value we utilize). To maintain connectivity in each direction (which is an important prerequisite in any NoC architecture), one subchannel is fixed in each direction, while the remaining subchannels can be reversed. This gives us three possible actions for four subchannels per RMC: (1,3), (2,2), and (3,1), where the numbers represent the number of subchannels in each direction. In other words:

$$A = \{a_1 = (1,3), a_2 = (2,2), a_3 = (3,1)\} \quad (2)$$

In this work, we consider four subchannels per RMC, however this formulation could easily be extended to any arbitrary number of subchannels (see Section 5.4 for a scalability analysis).

As mentioned in Section 3.3, in order for the RL policy to correctly pick the number of subchannels in each direction, the agent must be able to predict the future buffer demand. However, the buffer demand at a single RMC depends on the traffic in the surrounding network. Fortunately, we can leverage the information used in credit flow-control NoC architectures. In these architectures, each router contains a credit variable for each link that represents the number of free buffers in the next router. Every time a flit is transmitted along a link, its credit number is decremented, and every time a flit leaves the next router's input port, a credit is sent back up the link and the credit number is incremented. This provides us close to real-time information about the buffers in the region. Therefore, we define the state, $s_t$, to include the number of credits, $C_{ro}$, for each port from the RMC's two connecting routers, where $r$ refers to one of the two routers and $o$ refers to one of the four output ports. For example, the state for the RMC between routers 0 and 1 would be:

$$s = [C_{0N}, C_{0E}, C_{0S}, C_{0W}, C_{1N}, C_{1E}, C_{1S}, C_{1W}] \quad (3)$$

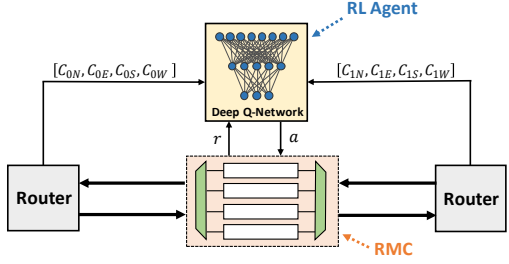

**Figure 3. The proposed RL framework for RMC control.**

While prior work only uses local information, e.g., QORE [6] uses link utilization, and CURE [7] uses buffer information (20-variable input state with information from the local link and router only), the credits stored in our two connected routers can provide us information about the routers *neighboring* the two connected routers. In this manner, our state captures a broader view of congestion than the local link and routers can provide.

In the end, the RMC controller should allocate subchannels in the direction that they would provide the most utility. To help measure this, we devise a new metric, "falsefull", which is the condition where all subchannels are full in one direction and there is at least one empty reversible subchannel in the other direction. In these cases, even though one direction sees a full RMC, the RMC is not really full and can be easily reconfigured to improve buffer utilization. Hence, we define the reward for each epoch as:

$$r_t = -\epsilon \cdot c_{unequal} - \sum_{t=0}^{T} f(t) \quad (4)$$

$$f(t) = \begin{cases} 1 & when\ falsefull\ exists \\ 0 & otherwise \end{cases} \quad (5)$$

$$c_{unequal} = \begin{cases} 1 & when\ a \neq (2:2) \\ 0 & otherwise \end{cases} \quad (6)$$

where $\epsilon$ is a small weighting factor to penalize non-equal subchannel actions, $c_{unequal}$ is the indicator function for the situation where the RMC action is *not* the (2,2) configuration (or equal subchannels), $T$ is the length of the epoch in cycles, and $f(t)$ is the indicator function for falsefull. We add a small additional cost for unequal allocations ((1,3) or (3,1)), $\epsilon \cdot c_{unequal}$, so that the agent is biased towards (2,2) when the falsefull cost is similar for different actions. As it is difficult to immediately reverse subchannels, this will help the RMC settle in the balanced position (2,2) when there is no advantage to the other actions, allowing it to more easily switch to (3,1) and (1,3). The value of $\epsilon$ can be adjusted to vary the importance of a balanced RMC in the policy, and is set to 1 in this work. With these state, action, and reward definitions (Fig. 3), we expect RACE to infer regional congestion and proactively configure RMCs to efficiently meet dynamic buffering demands across routers in a NoC.

## 5 RESULTS

### 5.1 Simulation and System Setup

For evaluating our proposed RACE framework, we consider a NoC with 64 nodes connected in an 8x8 mesh topology. We utilize the deadlock-free and low-overhead dimension order (XY) routing and credit-based flow control. For each input port of a router, we use 8 RMC buffers (along the connected link) and 2 router buffers (within the router), with each buffer being able to store a 128-bit flit. The 8 RMC buffers in each direction on a link (16 total) are organized into four subchannels with four RMC buffers each. As two subchannels are fixed in opposite directions, we can freely assign the other two subchannels in either direction. In this way, a port can be allocated three subchannels for a maximum of 14 (2 router + 3×4 RMC) buffers, and one subchannel for a minimum of 6 (2 router + 1×4 RMC) buffers depending on the decisions of the RMC control policy. Table 1 summarizes the network configuration details. In this work, we do not use virtual channels and consider it as one of our directions for future work.

**Table 1: NoC Platform Parameters**

| | |
|---|---|
| Number of cores | 64 |
| Topology | 8 × 8 mesh |
| Router buffers per port | 2 |
| RMC subchannels | 4 |
| RMC buffers per subchannel | 4 |
| Flow control | Credit-based |
| Link width | 128 bits |
| Routing scheme | Dimension-order (XY) |
| Operating frequency | 2 GHz |
| Operating voltage | 1 V |
| Process technology | 45 nm |

We compare our proposed RMC buffer control policy (RACE) against four alternatives: an approach with a static RMC allocation that has two subchannels fixed in both directions (FIXED); a heuristic-based RMC control policy, QORE [6]; an RL-based RMC control policy, CURE [7], and an "Oracle" policy that has two extra subchannels and always has three subchannels in each direction (ORACLE). The ORACLE policy represents the upper bound of performance that can be achieved by any RMC buffer control policy, while FIXED represents the performance if we do not adapt the RMC subchannels over time.

For power and performance evaluation, we modify the cycle-accurate network simulator Noxim [15] to simulate the network. We evaluate our platform using multi-application traffic. Here, we create eight multi-application benchmarks (s1 to s8), each consisting of a combination of eight different applications from the PARSEC benchmark suite [16], as shown in Table 2. The choice of eight applications per benchmark allows us to realize the type of traffic scenarios expected in a 64-core multiprocessor, where multiple multi-threaded applications would simultaneously execute across the available cores and utilize the NoC architecture. The total simulation time for all applications is set to 1,000,000 cycles, with a warm-up time of 100,000 cycles.

### 5.2 RMC Controller Setup

In QORE and RACE, every RMC has a dedicated controller to configure the subchannel directions. On the other hand, CURE

uses a router-based RMC controller which chooses a single configuration for all RMCs connected to the NoC router. Due to the router-based RMC controllers in CURE, each RMC will receive two configurations, one from each of their adjacent routers. These conflicts are resolved by prioritizing the configuration from the router south or east of the RMC. The control decisions are made every 50-cycle epoch for QORE and RACE and every 1000-cycle epoch for CURE (as advocated in [7]).

Both CURE and RACE RL agents are trained offline until convergence, with all agents (64 for CURE, 112 for RACE) interacting and learning independently from one another. For CURE, we train the agents for each benchmark set and evaluate it on that benchmark set (application-specific policies). Each CURE model uses a fully connected ANN consisting of 19 input neurons (originally 20, the temperature input was unrelated to the RMC control and hence is omitted in this work), 30 hidden layer neurons, and 3 output neurons. For RACE, we train the agents using s1-s5 as the training benchmark sets and evaluate them on all benchmark sets s1-s8 (note that s6-s8 are unseen by the agents during training). RACE's ANN has 8 input neurons, 5 hidden layer neurons, and 3 output neurons. We set the learning rate $\alpha$ to 0.001 and the discount factor $\gamma$ to 0.99.

The RMC controllers were implemented in RTL and synthesized in Cadence Genus for accurate power and performance estimates. We used a 45 nm technology library with a supply voltage of 1 V and an operating frequency of 2 GHz.

### 5.3 Performance, Power, and Area

Fig. 4 shows the normalized average network latency for the different RMC control policies with respect to FIXED. RACE is able to reduce the latency by 31%, 20.7%, and 48.9% on average compared to FIXED, QORE, and CURE, respectively. This is due to RACE's ability to correctly infer future congestion and adapt the RMC configuration to reduce falsefull events. As the ORACLE always assumes maximum subchannels in both directions (3, 3), it may be impossible to achieve that level of results. However, ORACLE provides a reference point for the upper bound in performance. It can be observed that our policy (RACE) is able to get very close to the ORACLE results in a few cases.

**Table 2: Multi-Application PARSEC Benchmark Sets**

| Benchmark set | Applications |
|---|---|
| s1 | flu.m, x264.s, ded.m, swa.l, bla.m, fer.m, flu.s, can.m |
| s2 | flu.m, x264.s, ded.m, swa.l, bla.m, fer.m, vip.m, can.m |
| s3 | flu.m, x264.s, ded.m, swa.l, bla.m, vip.m, flu.s, can.m |
| s4 | flu.m, x264.s, ded.m, swa.l, fer.m, vip.m, flu.s, can.m |
| s5 | flu.m, x264.s, swa.l, bla.m, flu.l, fer.m, vip.m, flu.s, can.m |
| s6 | x264.s, ded.m, swa.l, bla.m, fer.m, vip.m, flu.s, can.m |
| s7 | x264.s, ded.m, swa.l, bla.m, flu.l, fer.m, flu.s, can.m |
| s8 | x264.s, ded.m, swa.l, bla.m, flu.l, fer.m, vip.m, can.m |

**Table 3: Power Consumed for the RMC Approaches Compared to Baseline NoC with No RMCs**

| RMC Policy | FIXED | QORE | CURE | RACE |
|---|---|---|---|---|
| Power Consumption | 40.409% | 40.413% | 41.881% | 41.265% |

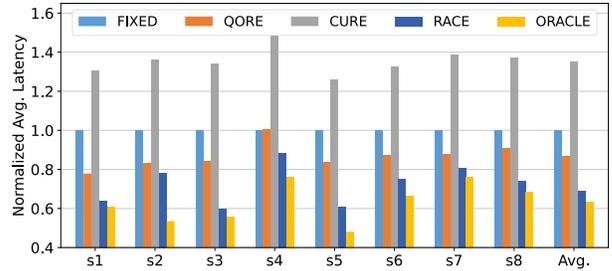

**Figure 4. Average network latency for the different RMC control policies normalized to FIXED.**

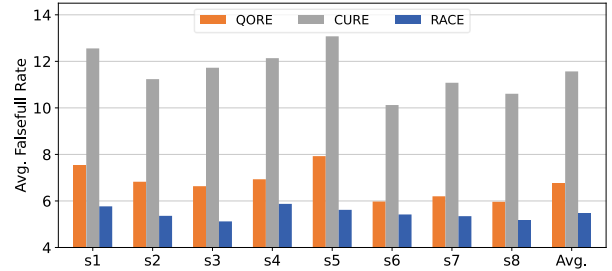

**Figure 5. The rate of falsefull (falsefulls per cycle) for the different RMC control policies.**

In Fig. 5, we show that the average number of falsefulls across the entire NoC for the three RMC control policies. Here, we see a similar trend to the latency results. RACE is able to reduce falsefulls more than QORE or CURE, indicating a better use of the RMC buffers in RACE. It should be noted that CURE unexpectedly does worse than even FIXED. To better understand the results for CURE, we look at the frequency of 1, 2, and 3 unique RMC configurations around each router in Fig. 6. Here, we can see that for the majority of the time (~75%), QORE and RACE have 2 or 3 unique RMC configurations around the router. As CURE sets one RMC configuration for all RMCs around the router, it cannot achieve the same granularity, leading to poor RMC buffer utilization. Due to its conflict resolution technique, CURE can sometimes have two unique RMC configurations around a router. Moreover, CURE uses a reconfiguration epoch of 1000 cycles which reduces its ability to quickly adapt to temporal changes in buffer demand.

Fig. 7 shows the energy consumed by the NoC for FIXED, QORE, CURE, and RACE normalized to FIXED. The NoC energy consumption for RACE is reduced by 34.7%, 14.6%, and 47.1% on average against FIXED, QORE, and CURE, respectively. Table 3 shows the power consumed by the NoC when using each type of RMC policy controller, as well as FIXED which does not need a controller, compared to a baseline NoC without RMCs and the same number of buffers all contained within the routers. It can be seen that RMCs lead to a substantial reduction in NoC power (about 40% lower power for FIXED vs. the baseline NoC). The RMC controllers in QORE, CURE, and RACE only raise the power consumption to 40.413%, 41.881%, and 41.265% respectively. The area overhead of the 112 RACE RL agents is 3.822 mm$^2$ whereas the 64 CURE RL agents take up 4.1 mm$^2$. Thus, our proposed RACE framework is able to provide significant improvement in network latency and energy consumption, with minimal power and area costs, compared to other RMC buffer control strategies.

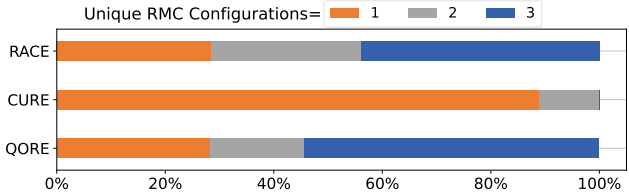

**Figure 6. The frequency of the number of unique RMC configurations around a NoC router.**

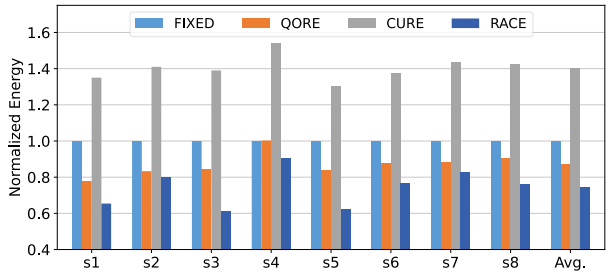

**Figure 7. NoC energy consumption for RMC control policies**

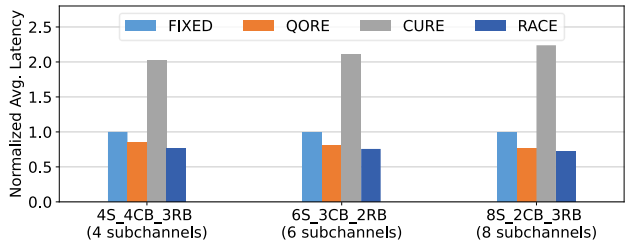

**Figure 8. Relative performance for the different policies with different number of RMC subchannels.**

### 5.4 Subchannel Sensitivity Analysis

Lastly, we tested the scalability of the various RMC buffer control policies by applying them to NoCs with more RMC subchannels. We adjust the number of router/RMC buffers to keep the same total buffers per port for all systems. For example, the RMC system (4S_4CB_3RB) denotes four subchannels, four channel buffers per subchannel, and three router buffers per port. Fig. 8 shows the average latency for each RMC control policy for 4, 6, and 8 RMC subchannels normalized to FIXED. The results are the average of all traffic sets (s1-s8). Here, we see very similar results to Fig. 4, and RACE is still the most effective in reducing network latency with increasing number of subchannels. In fact, as the control is more fine-grained for the greater number of subchannels, the policy actually improves performance by a greater margin. In contrast, CURE performs worse with increasing number of subchannels. As the number of subchannels increases, the number of possible unique configurations increases, only furthering the divide between RACE and CURE seen in Fig. 6.

### 6 CONCLUSION

In this paper, we proposed RACE, an RL-based control policy for NoC reversible channel buffers to improve NoC performance and energy-efficiency. RACE benefits from better awareness of network congestion and a novel reward metric ("falsefulls") to help guide the RL agent towards the best RMC control decisions. This, combined with a greater degree of reconfigurability and RL's proactiveness, allows RACE to learn effective and scalable policies for RMC control. These benefits result in RACE improving latency by 48.9%, and energy consumption by 47.1% on average against the state-of-the-art RL policy for RMC-based NoC platforms.

### REFERENCES

[1] Y. Hoskote, S. Vangal, A. Singh, N. Borkar, and S. Borkar, "A 5-GHz Mesh Interconnect for a Teraflops Processor," *IEEE Micro*, vol. 27, no. 5, Sept. 2007, pp. 51–61.

[2] B. K. Daya *et al.*, "SCORPIO: A 36-core research chip demonstrating snoopy coherence on a scalable mesh NoC with in-network ordering," in *Proc. Inter. Symp. on Comp. Arch.*, Jun. 2014, pp. 25–36.

[3] P. Kundu, "On-die interconnects for next generation CMPS," presented at *The Workshop on On- and Off-Chip Interconnection Networks for Multicore Syst.*, Stanford, CA, USA, Dec. 6–7, 2006

[4] A. K. Kodi, A. Sarathy, and A. Louri, "iDEAL: Inter-router Dual-Function Energy and Area-Efficient Links for Network-on-Chip (NoC) Architectures," in *Proc. Int. Symposium on Computer Arch. (ISCA)*, Jun. 2008, pp. 241–250.

[5] T. Moscibroda and O. Mutlu, "A case for bufferless routing in on-chip networks," in *Proc. Int. Symp. on Comp. Architecture*, Jun. 2009, pp. 196–207.

[6] D. DiTomaso, A. Kodi, and A. Louri, "QORE: A fault tolerant network-on-chip architecture with power-efficient quad-function channel (QFC) buffer," in *Proc. IEEE Int. Symp. on High Perform. Comp. Architecture*, Feb. 2014, pp. 320–331.

[7] K. Wang and A. Louri, "CURE: A High-Performance, Low-Power, and Reliable Network-on-Chip Design Using Reinforcement Learning," *IEEE Trans. Parallel Distrib. Syst.*, vol. 31, no. 9, Sep. 2020, pp. 2125–2138.

[8] T. Bjerregaard and S. Mahadevan, "A survey of research and practices of Network-on-chip," ACM Comput. Surv., vol. 38, no. 1, Jun. 2006, p. 1

[9] R. Hesse, J. Nicholls, and N. E. Jerger, "Fine-Grained Bandwidth Adaptivity in Networks-on-Chip Using Bidirectional Channels," in *Proc. IEEE/ACM Int. Symp. on Networks-on-Chip*, May 2012, pp. 132–141.

[10] D. DiTomaso, A. K. Kodi, A. Louri, and R. Bunescu, "Resilient and Power-Efficient Multi-Function Channel Buffers in Network-on-Chip Architectures," *IEEE Trans. Comput.*, vol. 64, no. 12, Dec. 2015, pp. 3555–3568.

[11] K. Wang, A. Louri, A. Karanth, and R. Bunescu, "IntelliNoC: A Holistic Design Framework for Energy-efficient and Reliable On-chip Communication for Manycores" in *Proc. Int. Symp. on Comp. Arch.*, Jun. 2019, pp. 589–600.

[12] Y. Li, "Deep Reinforcement Learning,", 2018, *arXiv:1810.06339*.

[13] V. Mnih *et al.*, "Playing Atari with Deep Reinforcement Learn.," 2013, *arXiv: 1312.5602*

[14] C. J. C. H. Watkins and P. Dayan, "Q-learning," *Machine Learning*, vol. 8, no. 3, May 1992, pp. 279–292.

[15] V. Catania, A. Mineo, S. Monteleone, M. Palesi, and D. Patti, "Noxim: An open, extensible and cycle-accurate network on chip simulator," in *Proc. IEEE Int. Conf. on Application-specific Syst., Architectures and Processors*, Jul. 2015, pp. 162–163.

[16] C. Bienia, "Benchmarking Modern Multiprocessors," Ph.D. dissertation, Dept. of Comp. Sci., Princeton University, NJ, Jan. 2011, p. 153.

[17] J.W. Dally and B. Towles, *Principles and Practices of Interconnection Networks*, San Francisco, CA, USA: Morgan Kaufmann Publishers, 2004.

[18] S. Pasricha, and N. Dutt, *"On-Chip Communication Architectures",* Burlington, NJ, USA: Morgan Kauffman, 2008.